\documentstyle[12pt,epsfig]{article}

\newcommand{\be}{\begin{equation}}
\newcommand{\ee}{\end{equation}}
\def\aprle{\buildrel < \over {_{\sim}}}
\def\aprge{\buildrel > \over {_{\sim}}}
\begin{document}  
\topmargin 0pt
\oddsidemargin=-0.4truecm
\evensidemargin=-0.4truecm
\renewcommand{\thefootnote}{\fnsymbol{footnote}}
\newpage
\setcounter{page}{0}
\begin{titlepage}   
\vspace*{-2.0cm}  
\begin{flushright}
FISIST/19-2002/CFIF \\
hep-ph/0209192
\end{flushright}
\vspace*{0.1cm}
\begin{center}
{\Large \bf
Solar neutrino oscillations 
\vspace*{0.16cm} 
and bounds on neutrino magnetic moment and solar magnetic field}

\vspace{0.6cm}

{\large 
E. Kh. Akhmedov\footnote{On leave from National Research Centre Kurchatov 
Institute, Moscow 123182, Russia. E-mail: akhmedov@cfif.ist.utl.pt} and
Jo\~{a}o Pulido\footnote{E-mail: pulido@cfif.ist.utl.pt}}\\  
\vspace{0.15cm}
{\em Centro de F\'\i sica das Interac\c c\~oes Fundamentais (CFIF)} \\
{\em Departamento de F\'\i sica, Instituto Superior T\'ecnico }\\
{\em Av. Rovisco Pais, P-1049-001 Lisboa, Portugal}\\  
\end{center}
\vglue 0.6truecm
\begin{abstract}
If the observed deficit of solar neutrinos is due to neutrino oscillations, 
neutrino conversions caused by the interaction of their transition magnetic 
moments with the solar magnetic field (spin-flavour precession) can still be 
present at a subdominant level. In that case, the combined action of neutrino 
oscillations and spin-flavour precession can lead to a small but observable 
flux of electron antineutrinos coming from the sun. Non-observation of these 
$\bar{\nu}_e$'s could set limits on neutrino transition moment $\mu$ and the 
strength and coordinate dependence of the solar magnetic field $B_\perp$. 
The sensitivity of the $\bar{\nu}_e$ flux to the product $\mu B_\perp$ is 
strongest in the case of the vacuum oscillation (VO) solution of the solar 
neutrino problem; in the case of the LOW solution, it is weaker,  and it 
is the weakest for the LMA solution.  For different solutions, different 
characteristics of the solar magnetic field $B_\perp(r)$ are probed: for 
the VO solution, the $\bar{\nu}_e$ flux is determined by the integral of  
$B_\perp(r)$ over the solar convective zone, for LMA it is determined by
the magnitude of $B_\perp$ in the neutrino production region, and for 
LOW it depends on the competition between this magnitude and the derivative 
of $B_\perp(r)$ at the surface of the sun. 
\end{abstract}
\vspace{.5cm}
\end{titlepage}   
\renewcommand{\thefootnote}{\arabic{footnote}}
\setcounter{footnote}{0}
\section{Introduction}
The observed deficit of solar neutrinos \cite{SNE} compared to the
expectations based on the standard solar model \cite{SSM} and the 
standard electroweak model \cite{SM} is now firmly established to be due
to non-standard neutrino properties. In particular, the SNO Collaboration has 
demonstrated \cite{SNO} that a significant fraction of solar $\nu_e$ 
is converted into some other active neutrino species, which can be $\nu_\mu$, 
$\nu_\tau$, $\bar{\nu}_\mu$ or $\bar{\nu}_\tau$. The most plausible and
widely accepted explanation of the observed solar neutrino deficit are
neutrino oscillations; 
however, some alternative possibilities are not ruled out yet. One of them
is neutrino spin-flavour precession \cite{SV} due to the interaction
of neutrino transition (flavour off-diagonal) magnetic moments with the
solar magnetic field. Unlike the ordinary neutrino spin precession
\cite{FS}, the spin-flavour precession (SFP) can take place even if
neutrinos are Majorana particles; in this case it converts left-handed
$\nu_e$ into right-handed $\bar{\nu}_\mu$ or $\bar{\nu}_\tau$, which would
be in accord with the SNO findings. Neutrino SFP can be resonantly enhanced 
in matter \cite{LM,Akh1}, very much similarly to the resonance
amplification of neutrino oscillations, the MSW effect \cite{MSW}.   
 
SFP of solar neutrinos, both resonance and non-resonance, can very well 
account for the observed solar neutrino deficit. It yields an excellent fit 
of all currently available solar neutrino data (see, e.g., [10 -- 18] for 
recent analyses), even somewhat better than that of the large mixing 
angle (LMA) oscillation solution, which is the best one among the oscillation 
solutions. 
However, to account for the solar neutrino data, SFP requires relatively
large values of the neutrino transition magnetic moment, $\mu \sim 10^{-11}
\mu_B$ for peak values of the solar magnetic field $B_0\sim 100$ kG. Although 
such values of $\mu$ are not experimentally excluded, they are hard to 
achieve in the simplest extensions of the standard electroweak model. 

In the present paper we shall be assuming that the neutrino transition 
magnetic moment and/or solar magnetic field strength are significantly below 
the values necessary for the SFP mechanism to account for the solar neutrino 
deficiency (though not completely negligible). Our assumption is that it
is neutrino oscillations that solve the solar neutrino problem, while the
SFP is present at a subdominant level. What can then the observable effects 
of SFP be? Its influence on the survival probability of solar $\nu_e$ will
be small and essentially indistinguishable from a small change of the 
neutrino oscillation parameters. However, the combined action of neutrino 
oscillations and SFP can lead to a qualitatively new effect which is absent 
when only oscillations or only SFP are operative  -- the production of a flux 
of electron antineutrinos \cite{LM,Akh2,next,Raju,APS,Baha}. 
Since all the currently favoured oscillation solutions of the solar neutrino 
problem -- LMA and LOW MSW solutions and vacuum oscillations (VO) -- require 
the solar neutrino oscillations to be driven by a large mixing angle 
\cite{osc},  an observable flux of solar $\bar{\nu}_e$ can in principle be 
produced. 

In  the present paper we address the question of what can be learned about 
the neutrino transition magnetic moments $\mu$ and the solar magnetic field 
by studying the solar $\bar{\nu}_e$'s. In particular, we 
discuss the bounds on $\mu$ and the strength and coordinate dependence of the 
solar magnetic field that can be derived from the current upper limits on
the solar $\bar{\nu}_e$ flux as well as from future experiments in case 
the flux of $\bar{\nu}_e$ from the sun is not observed. 

Experimentally, $\bar{\nu}_{e}$'s have a very clear signature and can be
easily distinguished from the other neutrino species. The main problem with
detecting $\bar{\nu}_{e}$'s from the sun is the background of electron
antineutrinos from nuclear reactors. This background is a steeply decreasing
function of neutrino energy; it becomes negligible for $E>$ (5 -- 8) MeV.
Therefore only the solar $^8$B neutrinos can contribute to an observable flux
of solar $\bar{\nu}_{e}$'s and the energy interval to be studied is 
$E\simeq$ (5 -- 15) MeV.

Neutrino magnetic moments can also manifest themselves through the additional
contribution to the $\nu e$ scattering cross section in the solar neutrino
detectors (see, e.g., \cite{nuscat} for recent discussions). However, these
contributions can only be noticeable if $\mu\aprge 10^{-10}\mu_B$. While
such large values of $\mu$ are consistent with the current laboratory upper
bounds \cite{lab}, they exceed the astrophysical bounds $\mu <$ (1 -- 3)   
$\times 10^{-12}\mu_B$ \cite{astro} by more than an order of magnitude. In
our study we shall be assuming the astrophysical upper bounds to be satisfied 
and so shall neglect the effects of neutrino magnetic moment on neutrino 
detection. 

\section{Probability of $\bar{\nu}_e$ production }

We shall assume neutrinos to be Majorana particles and consider their 
evolution under the combined action of oscillations and SFP. For simplicity, 
we shall discuss the case of just two neutrino flavours, $\nu_e$ and 
$\nu_\mu$, and their antineutrinos. 

There are essentially two ways in which $\bar{\nu}_e$'s can be produced:  
(1) the originally produced solar $\nu_e$ first oscillate into $\nu_\mu$,  
which are then converted into $\bar{\nu}_e$ by SFP; (2) solar 
$\nu_e$ first undergo SFP and get converted into $\bar{\nu}_\mu$, which 
then oscillate into $\bar{\nu}_e$. This can be schematically shown as  
\begin{eqnarray}
\nu_{eL} \buildrel \small{\rm osc.} \over {\longrightarrow} \nu_{\mu L}
\buildrel \small{\rm SFP} \over {\longrightarrow} \bar{\nu}_{eR}\,, \\
\nu_{eL} \buildrel \small{\rm SFP} \over {\longrightarrow} \bar{\nu}_{\mu R}
\buildrel \small{\rm osc.} \over {\longrightarrow} \bar{\nu}_{eR} \,.
\label{chains}
\end{eqnarray}
The oscillations and SFP in these two chains of conversions can either take 
place in the same spatial region, or be spatially separated. In the former 
case, the amplitudes of the processes (1) and (2) interfere. It was shown
in \cite{Akh2} that the interference is destructive, leading to a significant 
suppression of the solar $\bar{\nu}_e$ flux even if the probability of SFP 
is large. The reason for this is CPT invariance from which it follows that 
the matrix of the Majorana-type transition magnetic moments is 
antisymmetric; this, in turn, implies that the amplitudes (1) and (2) are
of opposite sign \cite{APS}. The cancellation between the amplitudes
(1) and (2) is exact when the corresponding intermediate states 
($\nu_\mu$ and $\bar{\nu}_\mu$) are degenerate, e.g., in vacuum; the
degeneracy is lifted by matter and/or twisting magnetic field (i.e. by a 
magnetic field whose direction in the plane transverse to the neutrino
momentum changes along the neutrino path) \cite{APS}, so inside the sun 
the $\bar{\nu}_e$ production is not completely blocked. Still, it is
strongly suppressed 
\footnote{Except, possibly, for the LOW solution of the solar
neutrino problem, see discussion in Sec. 3.2.}, 
and so we shall concentrate on the $\bar{\nu}_e$ production in the sequence 
of two spatially separated processes corresponding to eq. (2) \cite{Raju}:
first, SFP inside the sun converts solar $\nu_{eL}$ into $\bar{\nu}_{\mu R}$ 
which then oscillate into $\bar{\nu}_{eR}$ in vacuum on their way to the 
earth. The alternative possibility, corresponding to eq. (1), can be 
disregarded as the magnetic field in the region between the sun and the earth 
is negligibly small. The probability that a $\nu_{eL}$ born inside the sun 
will reach the earth as $\bar{\nu}_{eR}$ is then  
\begin{eqnarray} 
P(\nu_{eL}\to \bar{\nu}_{eR}) &=& P(\nu_{eL}\to\bar{\nu}_{\mu R}\,;\,R_\odot)
\cdot P(\bar{\nu}_{\mu R}\to\bar{\nu}_{eR}\,;\,R_{es}) \nonumber \\
&=& P(\nu_{eL}\to\bar{\nu}_{\mu R}\,;\,R_\odot)\cdot
\sin^2 2\theta \sin^2\left(\frac{\Delta m^2}{4E} R_{es}\right)\,,
\label{prob1}
\end{eqnarray}
where $R_\odot$ is the solar radius, $R_{es}$ is the distance between
the sun and the earth, $\Delta m^2$ and $\theta$ are the neutrino mass
squared difference and mixing angle, and $E$ is neutrino energy. We shall
now concentrate on the calculation of the SFP probability 
$P(\nu_{eL}\to\bar{\nu}_{\mu R}\,;\,R_\odot)$. 

The evolution of the neutrino system under the consideration is described 
by the following system of equations \cite{LM,Akh2,APS}:
\begin{eqnarray} 
i\,\nu_{eL}'\, &=& \,(V_e-c_2\delta)\;\nu_{eL}+s_2\delta\;\nu_{\mu L}
+\mu B_\perp e^{i \phi}\;\bar{\nu}_{\mu R}\,, \\
i\,\bar{\nu}_{eR}'\, &=& -(V_e+c_2\delta)\;\bar{\nu}_{eR}-\mu B_\perp 
e^{-i \phi}\;\nu_{\mu L} +s_2\delta\;\bar{\nu}_{\mu R} \,, \\
i\,\nu_{\mu L}'\, &=& \, s_2\delta\;\nu_{eL}-\mu B_\perp e^{i\phi}\;
\bar{\nu}_{eR}+(V_\mu+c_2\delta)\;\nu_{\mu L}\,, \\
i\,\bar{\nu}_{\mu R}'\, &=& \, \mu B_\perp e^{-i \phi}\;\nu_{eL} 
+s_2\delta\;\bar{\nu}_{eR}-(V_\mu-c_2\delta)\;\bar{\nu}_{\mu R}\,. 
\end{eqnarray} 
Here $V_e=\sqrt{2}G_F[N_e(r)-N_n(r)/2]$ and $V_\mu=\sqrt{2}G_F[-N_n(r)/2]$ are 
the matter-induced potentials of electron and muon neutrinos, respectively, 
$N_e$ and $N_n$ being the electron and neutron number densities; 
s$_2=\sin 2\theta$, $c_2=\cos 2\theta$, $\delta=\Delta m^2/4E$.   
In eqs. (4)-(7) the angle $\phi$ defines the direction of the magnetic
field ${\bf{B}}_\perp(r)$ in the plane orthogonal to the neutrino momentum, 
$B_\perp(r)=|{\bf{B}}_\perp(r)|$.  In the case of twisting magnetic fields 
($\phi=\phi(r)$), the probability of $\bar{\nu}_{eR}$ production inside
the sun can be enhanced \cite{APS,Baha}; since little is known about the 
possible twist of the solar magnetic field, we set $\phi=0$ hereafter. This 
approximation is expected to reduce the calculated flux of solar 
$\bar{\nu}_{eR}$'s and therefore make our upper bounds on $\mu B_\perp$ more 
conservative. 

Since we are interested in the situation when the SFP probability is small, 
it can be calculated in perturbation theory. To find the $\bar{\nu}_{eR}$ 
amplitude in leading order in $\mu B_\perp$, one needs to use in eq. (7) 
the ``unperturbed'' by SFP value of the amplitude $\nu_{eL}$. This means 
that the $\mu B_\perp$ terms in eqs. (4) and (6) should be omitted. These
two equations then 
decouple from the rest of the system and reduce to the standard two-flavour 
evolution equations describing neutrino oscillations in matter. Next, we take 
into account that, due to the partial cancellation between the amplitudes of 
transitions (1) and (2) discussed above, one can neglect the $\bar{\nu}_{eR}$ 
amplitude inside the sun; we therefore omit eq. (5) as well as the 
$\bar{\nu}_{eR}$ term in eq. (7). From eq. (7) one then obtains, up to an 
irrelevant phase factor, the following expression for the amplitude of the
$\nu_{eL}\to \bar{\nu}_{\mu R}$ transition: 
\be
A(\nu_{eL}\to\bar{\nu}_{\mu R}\;\,R_\odot)=\int
_{r_i}^{R_\odot}\!\mu
B_\perp(r)\,\nu_{eL}(r)\,e^{-i\int_{r_i}^r(V_\mu-c_2\delta)dr'}\,dr\,.
\label{Amp1} 
\ee
Here $r_i$ is the coordinate of the point at which the $\nu_{eL}$ was 
produced in the sun, and the amplitude $\nu_{eL}(r)$ is determined solely by 
neutrino oscillations, i.e. has to be found from eqs.~(4) and (6) with $\mu 
B_\perp=0$. 

It is easy to check that for the LMA and LOW solutions of the solar neutrino
problem the solar $\nu_{eL}\leftrightarrow \nu_{\mu L}$ oscillations are 
adiabatic; one therefore can use the adiabatic approximation, which yields  
\be
\nu_{eL}(r)=\cos\theta(r_i)\cos\theta(r)e^{-i\int_{r_i}^r E_1 dr'}+
\sin\theta(r_i)\sin\theta(r)e^{-i \int_{r_i}^r E_2 dr'}\,.
\label{nue1}
\ee
Here 
\be
E_{1,2}(r)=\frac{V_e+V_\mu}{2}\mp \omega\,,\qquad\quad 
\omega=\sqrt{\left(\,\frac{V_e-V_\mu}{2}-c_2\delta\right)^2+(s_2\delta)^2}\,,
\label{E1E2}
\ee
and $\theta(r)$ is the mixing angle in matter at a point $r$, defined through 
\be
\cos 2\theta(r)=\frac{c_2\delta-\frac{V_e-V_\mu}{2}}{\omega}\,. 
\label{theta}
\ee
Substituting (\ref{nue1}) into (\ref{Amp1}) one finds 
\be
A(\nu_{eL}\to\bar{\nu}_{\mu R}\;\,R_\odot)=\int_{r_i}^{R_\odot}\!\mu
B_\perp(r)\,[\cos\theta(r_i)\cos\theta(r)e^{-ig_1(r)}+
\sin\theta(r_i)\sin\theta(r)e^{-i g_2(r)}]\, dr\,,
\label{Amp2} 
\ee
where
\be
g_{1,2}(r)=\int_{r_i}^r\!\left(\,\frac{V_e+3 V_\mu}{2}-c_2\delta \mp
\omega\right) dr'\,.
\label{g1g2}
\ee
This expression is relevant for the LMA and LOW solutions of the solar 
neutrino problem. 

In the case of the VO solution, neutrino oscillations inside the sun are 
essentially blocked, so that instead of eq. (\ref{nue1}) one finds from 
eq. (4)   
\be
\nu_{eL}(r)=e^{-i\int_{r_i}^r (V_e-c_2\delta)\,dr'}\,.
\label{nue2}
\ee
Eq. (\ref{Amp1}) then yields
\be
A(\nu_{eL}\to\bar{\nu}_{\mu R}\;\,R_\odot)=\int_{r_i}^{R_\odot}\!\mu
B_\perp(r)\,e^{-i\int_{r_i}^r(V_e+V_\mu-2 c_2\delta)\,dr'}dr\,.
\label{Amp3} 
\ee

\section{Calculations of the expected $\bar{\nu}_{eR}$ flux}

Since only the solar $^8$B neutrinos with the energies $E>5$ MeV can 
contribute to an observable flux of solar antineutrinos, the expected flux 
of $\bar{\nu}_{eR}$'s in this energy region is $\Phi_{\bar{\nu}_e}(E)= 
\Phi_{^8\!B}(E)P(\nu_{eL}\to\bar{\nu}_{eR},E)$. We shall now concentrate on 
the calculation of the transition probability $P(\nu_{eL}\to\bar{\nu}_{eR},E)$ 
for the LMA, LOW and VO solutions of the solar neutrino problem. 

In eqs. (\ref{Amp2}) and (\ref{Amp3}), the pre-exponential factors in the 
integrands are in general smooth functions of $r$, but the complex 
exponential factors are very rapidly oscillating functions.
The oscillations are especially fast for the LMA solution because it 
corresponds to the largest values of $\Delta m^2$. The integrals of this 
type are notoriously difficult to calculate numerically -- one needs very 
fine integration steps in order for the integrals to converge. For 
example, for the LMA case a step $\sim10^{-6}R_\odot$ or smaller is 
necessary. 

Fortunately, there are well developed and accurate approximate analytic 
methods of calculating such integrals (see, e.g., \cite{Erdelyi}). For our 
purposes, analytic expressions also have an advantage of allowing one to 
directly relate the expected flux of solar $\bar{\nu}_{eR}$ to simple 
characteristics of the solar magnetic field.  
The integrals in eqs. (\ref{Amp2}) and (\ref{Amp3}) are of 
general type 
\be
I=\int_a^b f(x) e^{-i g(x)}\, dx\,,
\label{gen}
\ee
where $f(x)$ is a smooth function of $x$ and $|g'(x)|$ is large except 
possibly in the vicinity of a finite number of points in the interval 
$(a, b)$. The integrals of rapidly oscillating functions are in general 
strongly suppressed because the contributions of neighbouring points tend 
to cancel each other. The exceptions are the endpoints of the integration 
interval, for which there are no neighbouring points on one of the sides, 
and the points where $g'(x)=0$, which correspond to the extrema of the 
phase $g(x)$. In the vicinity of these (stationary phase) points the phase 
changes slowly and the corresponding contributions to the integral are not 
suppressed. These contributions can be found in the stationary phase 
approximation, which is the complex version of the steepest descent 
approximation. 

The contributions of the stationary phase points to the integrals of the 
type (\ref{gen}) are ${\cal O}(1/\sqrt{|g''|})$, whereas the endpoint  
contributions are in general ${\cal O}(1/|g'|)$. It is easy to check 
that for the integrals in eqs. (\ref{Amp2}) and (\ref{Amp3}) and for the 
values of neutrino parameters relevant for the LMA, LOW and VO solution of 
the solar neutrino problem the condition 
\be
|g''(x)|/g'(x)^2 \ll 1
\label{cond1} 
\ee
is always satisfied. Therefore, if there are stationary phase points in 
the integration interval, they are in general expected to give the dominant 
contributions to the integrals. However, as we shall show now, this 
does not happen in the cases of interest to us because the stationary
points either do not exist or their contributions are strongly suppressed 
by nearly vanishing pre-exponential factors in (\ref{gen}). 

Let us check the stationary phase conditions for various solutions of 
the solar neutrino problem. If one disregards neutrino oscillations and 
considers only SFP, the amplitude of the $\nu_{eL}\to \bar{\nu}_{\mu R}$ 
transitions is given by eq. (\ref{Amp3}). The stationary phase condition 
$g'=0$ then reduces to 
\be
V_e+V_\mu-2 c_2\delta=0\,.
\label{res1}
\ee
This is nothing but the resonance condition for neutrino SFP in the small 
$\theta$ limit \cite{LM,Akh1}.  As was discussed above, neutrino oscillations 
inside the sun can only be neglected in the case of the VO solution. Since 
for this solution the parameter $\delta$ is very small ($\sim 10^{-17}$ 
eV), the resonance condition (\ref{res1}) is satisfied essentially at the 
surface of the sun, where the magnetic field strength is known to be very 
small ($\sim$ 10 -- 100 G); therefore the stationary phase point plays no
role in this case.

For the LMA and LOW solutions, the stationary phase conditions $g_{1,2}'=0$ 
reduce to 
\be
V_e+V_\mu-2 c_2\delta=\frac{(s_2 \delta)^2}{2 V_\mu}\,,
\label{res2}
\ee
which is the SFP resonance condition in the presence of mixing. 
It is easy to see that this condition cannot be satisfied for large values 
of mixing angles $\theta$. Indeed, eq. (\ref{res2}) has solutions only 
when 
\be
\sin^2 2\theta \le \frac{1-Y_e}{Y_e}\,,
\label{cond2}
\ee
where $Y_e$ is the number of electrons per nucleon in matter. Its value 
varies  between 0.667 and 0.868 inside the sun, so that condition 
(\ref{cond2}) requires $\sin^2 2\theta < 0.5$, too small for both LMA and LOW 
solutions. Thus, there are no stationary phase points contributions to the 
amplitude of the $\nu_{eL}\to\bar{\nu}_{\mu R}$ transition in these cases 
as well, and one has to consider the endpoint contributions.

To obtain the contributions of the endpoints of the integration interval 
to an integral of the type (\ref{gen}) we integrate it by parts. 
Integrating two times one finds 
\be
\int_a^b f(x) e^{-i g(x)}\, dx=\left.\left[\left(i \frac{f(x)}{g'(x)} +
\frac{f'(x)}{g'(x)^2} - \frac{f(x) g''(x)}{g'(x)^3}\right)\!e^{-ig(x)}\right]  
\right|_a^b +{\cal O}(1/g'(x)^3)\,.
\label{gen2}
\ee
{}From (\ref{cond1}) it follows that the third term in the brackets can 
always be neglected. We shall now consider the amplitudes of the $\nu_{eL}\to
\bar{\nu}_{eR}$ transition for all three solutions of the solar neutrino 
problem of interest.

\subsection{LMA solution} 

The current best fit values of the neutrino parameters for this solution 
are $\Delta m^2\simeq 6\times 10^{-5}$ eV$^2$, $\sin 2\theta\simeq 0.9$
\cite{osc}. We will be interested in neutrino energies $E\simeq$ (5 -- 15)
MeV, so that $\delta\simeq$ (1 -- 3)$\times 10^{-12}$ eV. It is easy to check 
that the contribution of the first term in the square brackets in 
eq.~(\ref{Amp2}) is always at least one order of magnitude smaller than
that of the second term, and so we shall neglect it. We now apply formula 
(\ref{gen2}). One has $f'\propto (d/dr)[\sin\theta B_\perp]\simeq 
\sin\theta B_\perp'$ since in the adiabatic regime $\theta'$ is small. The 
comparison of the first and the second terms in the brackets in 
eq.~(\ref{gen2}) then shows that the first term dominates when 
\be
B_\perp(r_i)\gg |B_\perp'(r_i)| /g_2'(r_i)\simeq 10^{-4}\,R_\odot\,
|B_\perp'(r_i)|\,. 
\label{cond3}
\ee
We consider only the contribution of the neutrino production point $r_i$ 
since the magnetic field at the final point of evolution $r=R_\odot$ is 
negligible. Let us introduce the scale height for the solar magnetic field 
strength $L_B=|B_\perp^{-1}(dB_\perp(r)/dr)|^{-1}$, which is a characteristic 
distance over which the magnetic field varies significantly. Condition 
(\ref{cond3}) then can be written as 
\be
L_B(r_i) \gg 10^{-4}\,R_\odot\,.
\label{cond4}
\ee
Magnetic fields of scale heights as small as $L_B\aprle 10^{-4} R_\odot$
can only exist over very short distances and so cannot lead to any sizeable 
SFP.  We shall therefore consider large-scale solar magnetic fields which 
satisfy  (\ref{cond4}). 

\begin{figure}[h]
\setlength{\unitlength}{1cm}
\begin{center}
\epsfig{file=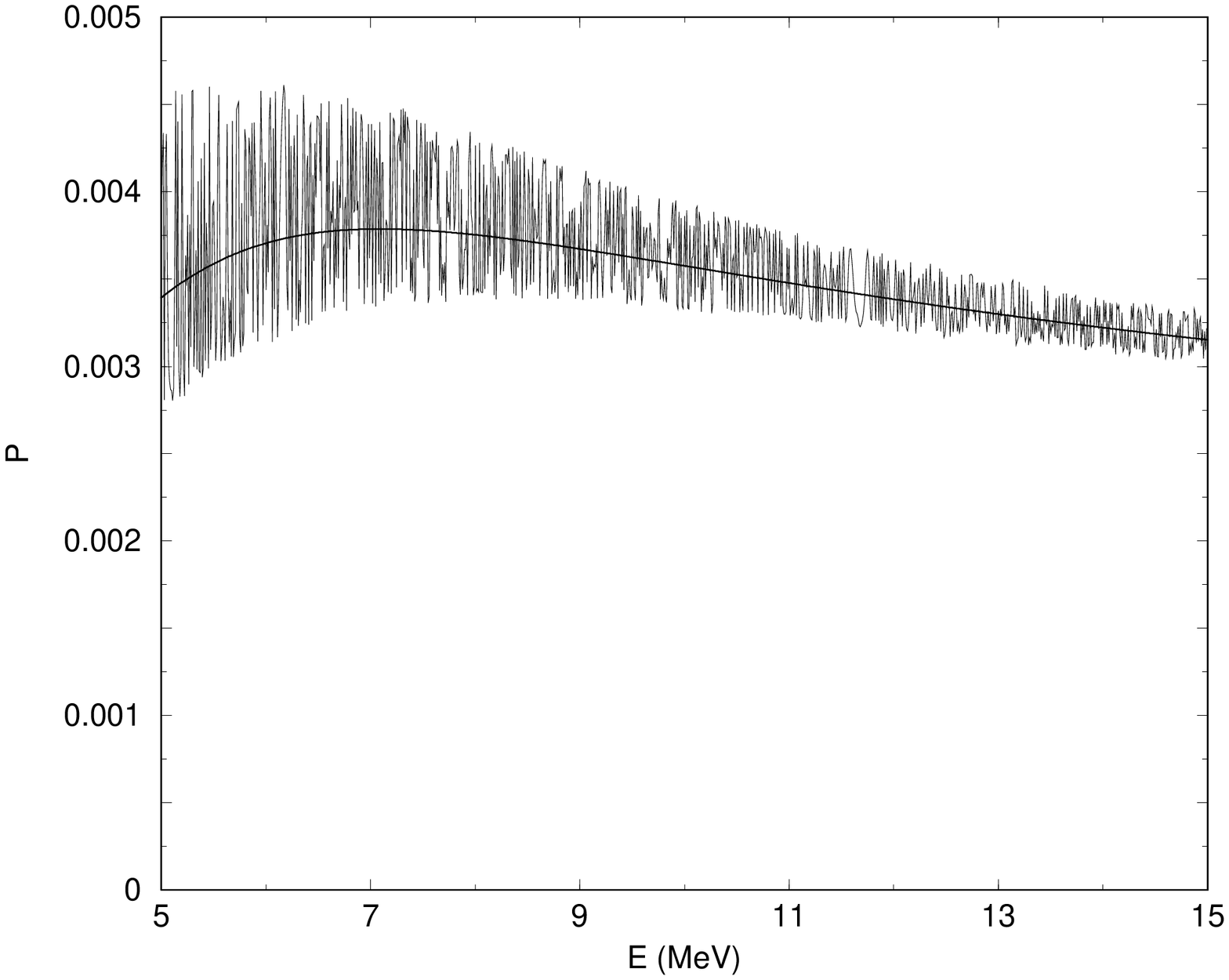,height=10.0cm,angle=270}
\end{center}
\vspace{-0.2cm}
{\small Figure 1: 
Probabilities $P(\nu_{eL}\to \bar{\nu}_{eR})$ corresponding
to the $\nu_{eL}\to \bar{\nu}_{\mu R}$ amplitudes of eqs. (\ref{Amp2}) (wiggly
curve) and (\ref{Amp4}) (smooth curve) for the LMA solution. Magnetic field
linearly decreasing from $B_0=5\times 10^7$ G at $r=0.05R_\odot$ to 
zero at $r=R_\odot$ and  $\mu=10^{-12}\mu_B$ were chosen. }
\end{figure}

The maximum of production of solar $^8$B neutrinos which are 
of interest to us corresponds to $r_0 \simeq 0.05 R_\odot$; from eqs. 
(\ref{gen2}) and (\ref{Amp2}) we then obtain 
\be
A(\nu_{eL}\to\bar{\nu}_{\mu R}\;\,R_\odot) \simeq \left.\left[\frac{\sin^2 
\theta(r_i) \mu B_\perp(r_i)}{g_2'(r_i)}\right]\right|_{r_i=0.05 
R_\odot} \,,
\label{Amp4} 
\ee
where we once again omitted an irrelevant phase factor. In the energy 
region of interest, the corresponding probability 
varies by less than 20\% (see fig. 1), and so in first approximation we
can replace it by its mean value.  From eqs. (\ref{prob1}), (\ref{Amp4})
and (\ref{g1g2}) one then finds 
\be
P(\nu_{eL}\to \bar{\nu}_{eR}) \simeq 1.8\times 10^{-10} 
\sin^2 2\theta \left[\frac{\mu}{10^{-12} \mu_B}
\frac{B_\perp(0.05R_\odot)}{10\,\mbox{kG}}\right]^2 \,,
\label{prob2}
\ee
where we have taken into account that the $\bar{\nu}_{\mu R}\to 
\bar{\nu}_{eR}$ oscillations in the space between the sun and the earth are 
in the averaging regime. Eq. (\ref{prob2}) is our final result for the 
LMA case.    

In fig. 1. the probabilities $P(\nu_{eL}\to \bar{\nu}_{eR})$ corresponding 
to the $\nu_{eL}\to \bar{\nu}_{\mu R}$ amplitudes of eqs. (\ref{Amp2}) and 
(\ref{Amp4}) are shown (the wiggly and smooth curves, respectively). 
The fast oscillations of the former are due to the interference between 
the two terms in eq. (\ref{Amp2}), whereas the latter is smooth because in 
obtaining it we neglected the (subleading) first term in (\ref{Amp2}). Notice 
that the oscillations described by the wiggly curve are in fact unobservable 
since they average out when one integrates over the neutrino production 
region or takes into account finite energy resolution of neutrino detectors. 

\subsection{LOW solution}
The best fit values of the neutrino parameters for this solution are
$\Delta m^2\simeq 10^{-7}$ eV$^2$, $\sin 2\theta\simeq 0.98$ \cite{osc}.
For the interval of neutrino energies of interest, the parameter $\delta$ 
is in the range $\delta\simeq$ (1.7 -- 5)$\times 10^{-15}$ eV. Once again, 
the contribution of the first term in the square brackets in eq. (\ref{Amp2}) 
is much smaller than that of the second term, and so we neglect it.
The analysis is similar to that in the LMA case, but there is one important 
difference. In the LMA case, the value of $g_2'$ changes only by about a 
factor of two from the neutrino production point to the surface of the sun
($g_2' R_\odot$ varies between $\sim 10^4$ and $\sim 5\times 10^3$). 
In contrast to this, in the LOW case it changes by a large factor -- from 
$g_2'R_\odot \sim 10^4$ at $r\simeq 0.05 R_\odot$ to $g_2'R_\odot \sim 10$ at 
$r=R_\odot$. This difference in the behaviour of $g_2'$ is a consequence of 
the fact that the values of $\delta$ are much larger and so contribute 
significantly to $g_2'$ at all values of $r$ in the LMA case whereas in the 
LOW case they dominate near the surface of the sun but are negligible in the 
solar core; the values of $g_2'$ at small $r$ are mainly determined by the 
matter-induced potentials. Indeed, in the LOW case one finds from eqs. 
(\ref{g1g2}) and (\ref{E1E2})
\be
g_2'|_{r=R_\odot}\simeq (1-c_2)\,\delta\,, \qquad\qquad 
g_2'|_{r=0.05R_\odot}\simeq (V_e+V_\mu)|_{r=0.05R_\odot} \,.
\label{approx}
\ee
The smallness of $g_2'$ at $r=R_\odot$ implies that the integral in eq. 
(\ref{Amp2}) is now dominated by the surface of the sun. Since the magnetic 
field nearly vanishes there, one can expect the main contribution to the 
integral to come from the second term in brackets in eq. (\ref{gen2}), i.e. 
from the term  proportional to the derivative of the solar magnetic field 
rather than to the field itself. Indeed, the condition for the domination of 
the term $\propto B_\perp'$ is  
\be
B_\perp(R_\odot)\ll [|B_\perp'(r)|/g_2'(r)]|_{r=R_\odot}\,,\quad
\mbox{or} \, \quad L_B(R_\odot)\ll 0.1\,R_\odot\,.
\label{cond5}
\ee
Since the magnetic field at the surface of the sun is very weak, this
condition is likely to be satisfied. 
For example, it is satisfied for any magnetic field profile which near the 
surface of the sun has the form $B_\perp=B_0[1-(r/R_\odot)^n]+B_1$, provided 
that $n>10(B_1/B_0)$; for the magnetic field at the solar surface $B_1=100$ G 
and the peak value of the  field $B_0=10$ kG this requires $n>0.1$. In what 
follows we shall be assuming that condition (\ref{cond5}) is satisfied.

The amplitude of the $\nu_{eL}\to\bar{\nu}_{\mu R}$ transition is then
approximately given by 
\be
A(\nu_{eL}\to\bar{\nu}_{\mu R}\;\,R_\odot) \simeq\sin\theta(r_i)\sin\theta\;
\frac{\mu \,|B_\perp(r)'|_{r=R_\odot}}{(1-c_2)^2 \,\delta^2}\,.
\label{Amp5} 
\ee
We shall parameterize the derivative of the magnetic field strength at 
$r=R_\odot$ as 
\be
|B_\perp(r)'|_{r=R_\odot}=\frac{B_0}{0.15\,R_\odot}\,\kappa\,,
\label{param}
\ee
with $B_0$ the peak value of the field of the convective zone ($0.7R_\odot\le 
r \le R_\odot$), and $\kappa$ a parameter; $\kappa=1$ would correspond to, 
e.g., a linearly decreasing towards the surface of the sun field with the 
maximum at the center of the convective zone. The probability of the 
$\nu_{eL}\to \bar{\nu}_{eR}$ transition is then
\be
P(\nu_{eL}\to \bar{\nu}_{eR}) \simeq 1.85\times 10^{-5} \,
\frac{\cos^2\theta}{\sin^4\theta}\,\kappa^2 \left(\frac{3\times  
10^{-15}\;\mbox{eV}}{\delta}\right)^4 \left[\frac{\mu}{10^{-12} 
\mu_B}\frac{B_0}{10\,\mbox{kG}}\right]^2 \,,
\label{prob3}
\ee
where we have taken into account that the $\bar{\nu}_{\mu R}\to 
\bar{\nu}_{eR}$ oscillations in the space between the sun and the earth are 
in the averaging regime, and that the mixing angle in matter at the 
neutrino production point $\theta(r_i)\simeq \pi/2$. 
As an example, for the best fit values of neutrino parameters for the LOW 
solution, $E=10$ MeV, $\mu=10^{-12}\mu_B$, $B_0=100$ kG and $\kappa=1$, 
eq.~(\ref{prob3}) yields $P(\nu_{eL}\to \bar{\nu}_{eR}) \simeq 1.4\times 
10^{-2}$.

\begin{figure}[h]
\setlength{\unitlength}{1cm}
\begin{center}
\epsfig{file=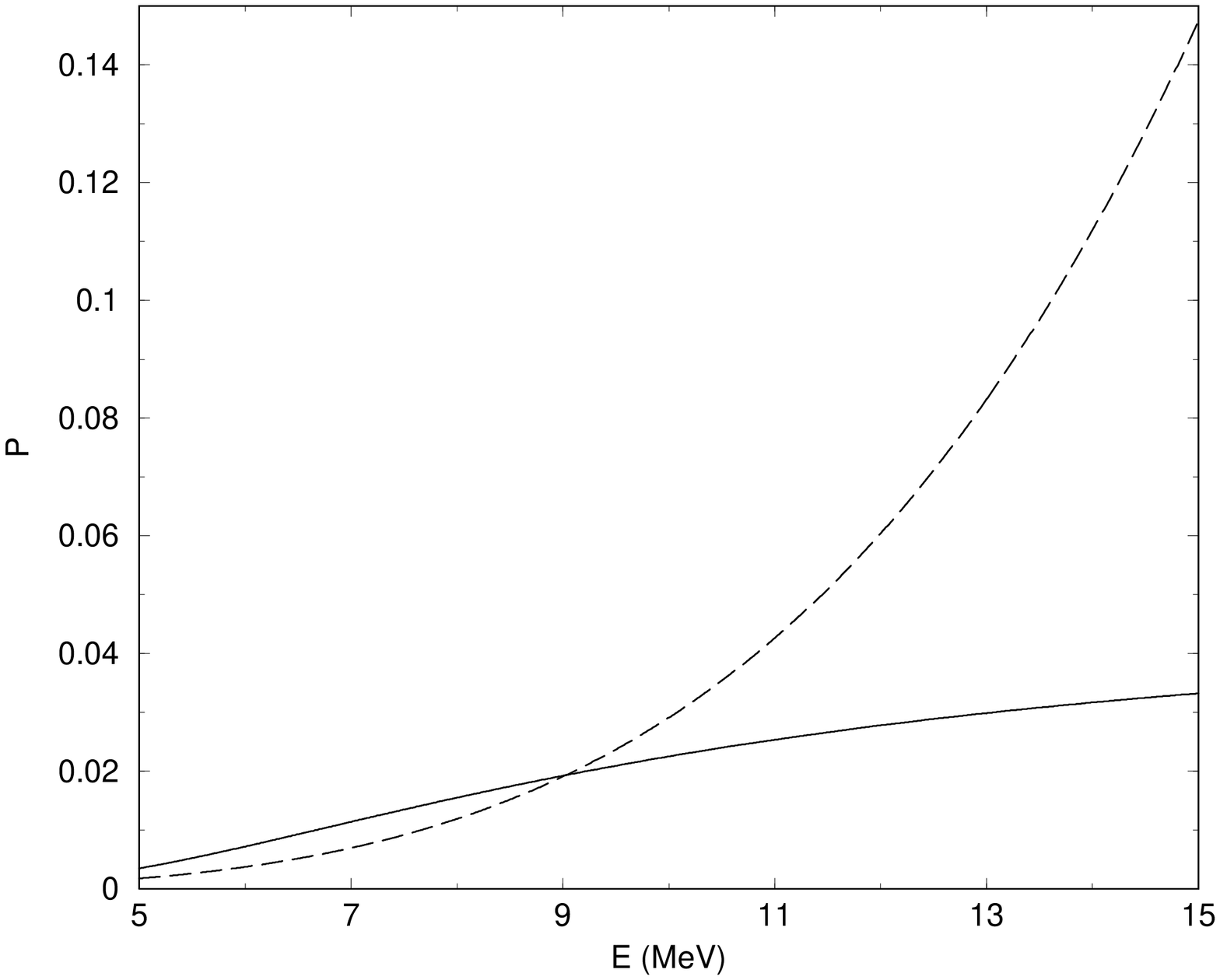,height=10.0cm,angle=270}
\end{center}
\vspace{-0.2cm}   
{\small Figure 2: Probabilities $P(\nu_{eL}\to \bar{\nu}_{eR})$ 
corresponding to the $\nu_{eL}\to \bar{\nu}_{\mu R}$ amplitudes of eqs. 
(\ref{Amp2}) (solid curve) and (\ref{Amp5}) (dashed curve) for the LOW 
solution. Magnetic field decreases linearly from $B_0=10^6$ G at 
$r=0.05R_\odot$ to zero at $r=R_\odot$, $\mu=10^{-12}\mu_B$. }
\end{figure}

It should be noted that, while the probability $P(\nu_{eL}\to\bar{\nu}_{eR})$ 
based on the perturbation-theoretic $\nu_{eL}\to \bar{\nu}_{\mu R}$ amplitude 
of eq. (\ref{Amp2}) (fig. 2, solid curve) is very accurate, the simplified 
expression (\ref{prob3}) (dashed curve in fig. 2) is only correct within a 
factor of three or four. The reason for that is the 
following. In the case of the LOW solution the quantity $g_2'R_\odot$
changes only by about a factor of two throughout the solar convective zone, 
being $\sim $ 10 -- 20, i.e. not too large. Therefore the contribution of 
the integration endpoint at the solar surface to the integral in (\ref{Amp2}) 
is not much bigger than that of the convective zone, and the approximation
(\ref{prob3}) is only an order of magnitude estimate. 

In deriving eqs. (\ref{Amp5}) and (\ref{prob3}) we have assumed, in addition 
to (\ref{cond5}), that the contribution of the term proportional to the 
magnetic field at the neutrino production point $r_i$  can be neglected
compared to the contribution of the term proportional to the derivative of 
the field at the solar surface. The condition for this is 
\be
B_\perp(r_i)<\frac{(V_e+V_\mu)|_{r_i}}{(1-c_2)^2\delta^2}\,
|B_\perp'|_{r=R\odot}\,, \qquad \mbox{or} \qquad 
B_\perp(r_i)\aprle 10^{3}\,B_0\,\kappa\,.
\label{newcond}
\ee
If it is not satisfied, 
the probability $P(\nu_{eL}\to \bar{\nu}_{eR})$ in the LOW case will be 
approximately given by eq. (\ref{prob2}), as it does in the LMA case. 

We pointed out earlier that the production of $\bar{\nu}_{eR}$ inside
the sun is suppressed because of the partial cancellation between the 
amplitudes of the channels (1) and (2). This cancellation is, however, 
partly compensated in the case of the LOW solution by an enhancement due to 
the fact that $c_2\delta$ is small and so is $V_e$ in the solar
convective zone, which is the region where the SFP mainly occurs. 
Therefore in the case of the LOW solution there is an additional channel of 
the $\bar{\nu}_{eR}$ production -- direct production inside the sun through 
neutrino oscillations and SFP. At the same time, the flux of $\bar{\nu}_{eR}$'s 
generated via the mechanism that we discussed so far is somewhat reduced if 
one takes the direct production into account; as a result, the total 
$\bar{\nu}_{eR}$ flux changes only slightly.  We have checked by a numerical 
integration of the system of the differential equations (1)-(4) that our 
approximation, in which we neglect the direct production but consider 
unsuppressed  production through the chain of the processes (2) yields a 
very accurate prediction of the total solar $\bar{\nu}_{eR}$ flux. 

\subsection{VO solution}

The best fit values of the neutrino parameters for this solution are
$\Delta m^2\simeq 4.5\times 10^{-10}$ eV$^2$, $\sin 2\theta\simeq0.93$ 
\cite{osc}.
The $\nu_{eL}\to \bar{\nu}_{\mu R}$ transition amplitude is given by 
eq. (\ref{Amp3}). For the interval of neutrino energies of interest, 
the parameter $\delta$ is in the range $\delta\simeq (0.77 - 2.3)\times 
10^{-17}$ eV, i.e. is negligible compared to $V_e+V_\mu$ essentially 
everywhere in the sun. One can therefore put $\delta=0$ in eq. (\ref{Amp3}), 
i.e. the $\nu_{eL}\to \bar{\nu}_{\mu R}$ transition probability is 
practically energy independent. 

In the case of the VO solution,  the parameter $g'$ that plays a key role
for integrals of rapidly oscillating functions of the type (\ref{gen}) 
is simply $g'=V_e+V_\mu$. The quantity $g'(r)R_\odot$ changes from a value 
$\sim 10^4$ in the center of the sun to nearly zero at its surface, being
$\aprle 1$ inside the convective zone. Thus, it is not legitimate to 
use the approximate expressions of the type (\ref{gen2}) which require 
$g'R_\odot$ to be very large everywhere in the integration interval. 
We shall therefore estimate the transition amplitude differently in this 
case.  

Let us first notice that the phase $g(r)$ that corresponds to 
eq. (\ref{Amp3}), 
\be 
g(r)=\int_0^r(V_e+V_\mu)\,dr'\,,
\label{phase}
\ee
first grows rapidly with $r$ and then saturates and slowly approaches 
its asymptotic value because of the steep decrease of $V_e+V_\mu$. The
contribution to the integral (\ref{Amp3}) from the region where $g(r)$ 
rapidly grows is strongly suppressed, and the main contribution comes from 
the region where the phase changes little. We can therefore adopt an 
approximation of retaining only the contribution of the region $r>r_0$ where 
the change of the phase $\Delta g \aprle 1$ and neglecting the phase change 
in this interval. Since it is the change of the phase and not its absolute 
value that matters for the transition probability, this amounts to
merely replacing in this interval the integrand of eq. (\ref{Amp3}) by
$\mu B_\perp(r)$. The lower boundary of the new integration interval $r_0$ is 
defined from the condition 
\be
g(R_\odot)-g(r_0)\simeq 1\,.
\label{r0}
\ee
As a result, we obtain 
\be
P(\nu_{eL}\to\bar{\nu}_{eR})\simeq \left|\mu\int_{r_0}^{R_\odot}\!B_\perp(r)
\,dr \right|^2 \, \sin^2 2\theta \sin^2\left(\frac{\Delta m^2}{4E}
R_{es}\right)\,.
\label{prob4}
\ee
Solving eq. (\ref{r0}) for $r_0$ we find $r_0\simeq 0.817 R_\odot$. We 
compared the $\nu_{eL}\to\bar{\nu}_{eR}$ probabilities obtained using this 
approximation with those based on the full $\nu_{eL}\to \bar{\nu}_{\mu R}$ 
amplitude (\ref{Amp3}) for a number of magnetic field profiles and found that 
the accuracy of the approximation (\ref{prob4}) is typically about 20\%. 

Let us introduce the average magnetic field strength in the interval $r_0\le 
r \le R_\odot$ through 
\be
\overline{B}_\perp=\frac{1}{R_\odot-r_0}\int_{r_0}^{R_\odot}\!B_\perp(r)\,dr\,,
\label{aver}
\ee
where $R_\odot-r_0=0.183 R_\odot$. Then eq. (\ref{prob4}) can be rewritten as 
\be
P(\nu_{eL}\to\bar{\nu}_{eR})\simeq 1.4\times 10^{-3}\left[\frac{\mu}{10^{-12} 
\mu_B}\frac{\overline{B}_\perp}{10\,\mbox{kG}}\right]^2 \, 
\sin^2 2\theta \sin^2\left(\frac{\Delta m^2}{4E}
R_{es}\right)\,.
\label{prob5}
\ee

\section{Discussion}

We calculated the probability of production of solar $\bar{\nu}_{eR}$'s  
assuming that the solar neutrino deficit is due to neutrino oscillations 
while the spin-flavour precession caused by the interaction of neutrino 
transition magnetic moments with the solar magnetic field is present as a 
subdominant process. We considered the SFP in perturbation theory and 
obtained analytic expressions for the transition probability $P(\nu_{eL} 
\to\bar{\nu}_{eR})$ valid for the LMA, LOW and VO solutions of 
the solar neutrino problem. We compared these analytical expressions 
with the results of numerical integration of the system of differential 
equations (1)-(4) and found very good agreement in all the cases.

For each of the solutions of the solar neutrino problem we then obtained
simplified approximate expressions for $P(\nu_{eL}\to\bar{\nu}_{eR})$, which
allowed us to relate this probability with simple characteristics of the  
solar magnetic field $B_\perp(r)$. For different solutions, different
characteristics of the solar magnetic field $B_\perp(r)$ are probed: for
the VO solution, the $\bar{\nu}_e$ flux is determined by the integral of
$B_\perp(r)$ over the upper 2/3 of the solar convective zone, for LMA it is 
determined by the magnitude of $B_\perp$ in the neutrino production region, 
and for the LOW solution it depends on the competition between this
magnitude and the derivative of $B_\perp(r)$ at the surface of the sun.    

The accuracy of the simplified expressions for $P(\nu_{eL}\to\bar{\nu}_{eR})$ 
is also different for different solutions: the error is less than 3\% for the 
LMA solution and about 20\% for the VO solution, while for the LOW solution
the simplified expression is only correct within a factor of three or four 
\footnote{Assuming that condition (\ref{newcond}) is satisfied. Otherwise,  
the description of the LOW case is similar to that of LMA, and the accuracy 
of the approximate expression is as good as it is in the LMA case.}.
Since the efficiency of SFP depends on the product of neutrino magnetic moment
and magnetic field strength, only this product and not $\mu$ and $B_\perp$
separately can be probed by studying the solar $\bar{\nu}_e$ flux 
$\Phi_{\bar{\nu}_e}$. Comparing eqs. (\ref{prob2}), (\ref{prob3}) and
(\ref{prob5}) we find that the sensitivity of the $\bar{\nu}_e$ flux to the
product $\mu B_\perp$ is strongest in the case of the VO solution of the  solar
neutrino problem; it is weaker in the case of the LOW solution and weakest for
the LMA solution. This, however, does not necessarily mean that the LMA
solution has the lowest sensitivity to the neutrino magnetic moment: the   
$\bar{\nu}_{eR}$ flux in that case depends on the magnetic field in the core
of the sun which may well be much stronger than the field in the convective   
zone, relevant for the VO case.

We shall now discuss the present experimental upper bounds on
$\Phi_{\bar{\nu}_e}$ as well as the sensitivity of the future experiments, 
and their implications. Currently, the most stringent upper bounds on
$\Phi_{\bar{\nu}_e}$ come from the LSD experiment, $\Phi_{\bar{\nu}_e}<
(1.7\times 10^{-2})\,\Phi_{^8\!B}$ at 90\% CL \cite{LSD}, and from the 
Super-Kamiokande experiment, $\Phi_{\bar{\nu}_e}< (1.2 - 1.6) \times
10^{-2}\,\Phi_{^8\!B}$ at 90\% CL \cite{Smy}. The Super-Kamiokande bounds 
were presented for several energy bins with $E\ge 8$ MeV. Future experiments 
are expected to improve these bounds (or discover the flux of solar 
$\bar{\nu}_{eR}$): KamLAND will be able to put a limit of $10^{-3}\,
\Phi_{^8\!B}$ at 95\% CL on the solar $\bar{\nu}_{eR}$ flux after one year 
of operation \cite{KamLAND}, and Borexino should be able to reach a similar 
sensitivity after a few years of data taking \cite{Lothar}. The current 
limit $\Phi_{\bar{\nu}_e} \aprle 1.5\%\,\Phi_{^8\!B}$ implies, in the case of 
the LMA solution, a bound 
\be
\left[\frac{\mu}{10^{-12}\mu_B}\frac{B_\perp(r_i)}{10\,\mbox{kG}}\right]  
\aprle 10^4\,,
\label{bound1}
\ee
where $B_\perp(r_i)$ is the average solar magnetic field in the neutrino 
production region, $r\aprle 0.1R_\odot$. An experiment with the tritium
radioactive antineutrino source has been recently proposed with the goal of 
putting an upper limit of $3\times 10^{-12}\mu_B$ on the neutrino magnetic 
moment $\mu$ or measuring it if it is above this value \cite{source}; if 
$\mu\simeq 3\times 10^{-12}\mu_B$ is found, the bound (\ref{bound1}) would 
imply $B_\perp(r_i) \aprle 3\times 10^7$ G. Note that this limit is more 
stringent than the astrophysical one obtained from the requirement that the 
pressure of the solar magnetic field should not  exceed the matter pressure
($B_\perp \aprle 10^9$) \cite{astrolim}. 
   
Conversely, if a reliable quantitative model of the solar magnetic field is
developed, eq.~(\ref{bound1}) will limit the neutrino magnetic moment. For
$B_\perp(r_i)$ close to the above-mentioned astrophysical bound, the limit    
on $\mu$ would be $\mu \aprle 10^{-13}\mu_B$, which is more than an order
of magnitude more stringent than the expected limit from the planned 
laboratory experiment \cite{source}. Unfortunately, no compelling model of
the solar magnetic field exists at present.

Similar considerations apply to the LOW and VO cases. For LOW, assuming 
that condition (\ref{newcond}) is satisfied, the current limits on the
solar $\bar{\nu}_{eR}$ flux lead to 
\be
\kappa \left[\frac{\mu}{10^{-12}\mu_B}\frac{B_0}{10\,\mbox{kG}}
\right] \aprle 10\,,
\label{bound2} 
\ee
where $\kappa$ and $B_0$ parameterize the derivative of the solar magnetic
field at $r=R_\odot$, see eq.~(\ref{param}). For $\mu\simeq 3\times 10^{-12}
\mu_B$, eq. (\ref{bound2}) limits this derivative to be 
$|B_\perp'(r)|_{R_\odot}\aprle 2.2\times 10^2$ kG/$R_\odot$. Conversely,  
if the actual value of $|B_\perp'(r)|_{R_\odot}$ is close to this value,
the limit on $\mu$ from (\ref{bound2}) would be competitive with the expected
upper bound from the planned laboratory experiment \cite{source}. If condition 
(\ref{newcond}) is not satisfied, the preceding discussion of the LMA case   
applies to the LOW case as well. 

In the VO case, the current upper bounds on $\Phi_{\bar{\nu}_e}$ imply 
\be
\left[\frac{\mu}{10^{-12}\mu_B}\frac{\overline{B}}{10\,\mbox{kG}}
\right] \aprle 5\,,
\label{bound3} 
\ee
where $\overline{B}$ is the average magnetic field in the interval $0.817 
R_\odot\le r \le R_\odot$ defined in (\ref{aver}). For $\mu=3\times
10^{-12}$ this gives $\overline{B}<17$ kG. If, alternatively, some model
considerations establish that the average field $\overline{B}$ is, for 
example, 100 kG, eq. (\ref{bound3}) would lead to the limit 
$\mu < 5\times 10^{-13}$. 

With the expected upper bound on the flux of solar $\bar{\nu}_{eR}$ from
KamLAND, all the limits that we discussed above (eqs. (\ref{bound1}) 
-- (\ref{bound3})) will be strengthened by about a factor of four.

In the VO case, the flux of solar $\bar{\nu}_{eR}$'s will have seasonal 
variations, similar to those of the $\nu_{eL}$ flux. In the LOW and VO 
cases, for which the $\bar{\nu}_{eR}$ production is mainly driven by 
the magnetic field in the convective zone, the solar $\bar{\nu}_{eR}$ flux
may also vary with time due to the 11-year variations of this magnetic
field. For all the solutions, the solar $\bar{\nu}_{eR}$ flux should, of 
course, also exhibit $\sim 7\%$ variations due to the variations of the 
distance between the sun and the earth. Low statistics may, however, make 
these variations difficult to detect.

\vspace{0.1cm}
\noindent

{\em Acknowledgements.} We are grateful to A. Mour\~ao for useful discussions.
E.A. was supported by the Calouste Gulbenkian Foundation as a Gulbenkian 
Visiting Professor at Instituto Superior T\'ecnico.

\end{document}